# Preparation, characterization and X-ray photoemission spectroscopy study of a correlated semimetal, SmBi

Anup Pradhan Sakhya, Arindam Pramanik, Ram Prakash Pandeya, Sawani Datta, Abhishek Singh, A. Thamizhavel and Kalobaran Maiti [a)]

*Department of Condensed Matter Physics and Materials Science, Tata Institute of Fundamental Research, Homi Bhabha Road, Colaba, Mumbai - 400 005, INDIA.*

[a)]Corresponding author: kbmaiti@tifr.res.in

**Abstract.** We report high quality single crystalline growth of SmBi using flux method. The compound crystallizes in the simple rock salt structure with space group *Fm3m* (No. 225). The cubic structure of the single crystal was confirmed by Laue diffraction pattern. The magnetic susceptibility measurements revealed sharp antiferromagnetic order with Néel temperature, $T_N$ = 9K. The core-level photoemission spectroscopy study of Sm 3*d* has been performed using monochromatic Al *Kα* ($h\upsilon$ =1486.6 eV) laboratory source. We observe multiple features in the experimental spectra due to final state effect – a signature of hybridization between Sm 4*f* - Bi6*p* states. Intense satellite features are also observed presumably due to mixed valency arising from Kondo coupling. No signature of surface-bulk difference is observed in the Sm 3*d* core level spectra.

## INTRODUCTION

The discovery of topological insulators has generated enormous interest in condensed matter physics recently because they host new exotic quantum states of matter such as Dirac or Weyl fermions [1]. Experimentally, majority of the observed topological materials discovered till date are weakly correlated electron systems, where the electron correlation can be usually ignored. As of late, the correlated electron systems along with strong spin-orbit coupling (SOC) have been the subject of intensive research [2]. These materials can also exhibit exotic physics due to the interplay between electron correlation and topological protection leads to exotic behaviour in correlated systems [3,4]. One of the most discussed example of a correlated topological material is the Kondo insulator $SmB_6$, which was recently predicted to host exotic topological surface states [5-7], though there have been recent studies, which cast doubts on the insulating property of $SmB_6$ [8-12]. Recently, another Sm based material, SmBi belonging to the family of rare-earth mono-bismuthides have generated considerable research interest as potential candidates for correlated topological semimetals [13]. The material crystallizes in the simple rock salt structure (face-centered cubic) having the space group *Fm3m*. Here, we discuss the preparation of high quality single crystalline SmBi and its characterization by using various techniques. X-ray photoelectron-emission spectroscopy (XPS) has been performed to understand the electronic structure of this material. To find the difference or similarities between the electronic structure of the bulk and the surface, we have varied the electron emission angle.

## EXPERIMENT

Single crystal of SmBi was grown from Indium flux method with a molar ratio of Sm:Bi:In of 1:1:10. High purity Sm (99.9%), Bi (99.998%) and In (99.9999%) were weighed and put into $Al_2O_3$ crucible and subsequently sealed in an evacuated quartz ampoule at a vacuum of $10^{-6}$ Torr. The sealed ampoule was then placed in a box furnace, where the temperature was increased at a rate of 60 ºC/hr up to 1050 ºC and kept for 24 hr. After that the temperature was reduced very slowly at the rate of 2 ºC/hr to 700 ºC, and at that temperature, the ampoule was centrifuged to remove the In flux. After this process, we were able to get shiny single crystals of SmBi. The typical size of the crystal is in mm range. The composition and purity of the synthesized crystal were determined by the spectra obtained by energy dispersive analysis of *x*-rays, which shows both Sm and Bi in the ratio of approximately 1:1. The crystal structure and the Laue diffraction pattern of the SmBi crystal is shown in Fig. 1. The Laue diffraction pattern observed experimentally was similar to the pattern observed from the Laue XRD simulation using software Orient Express V3.3 which indicates good crystallinity of the sample and the surface shown here corresponds to the (001) plane of the crystal. Well defined spots with clear symmetry pattern confirm that the grown single crystals are of good quality.

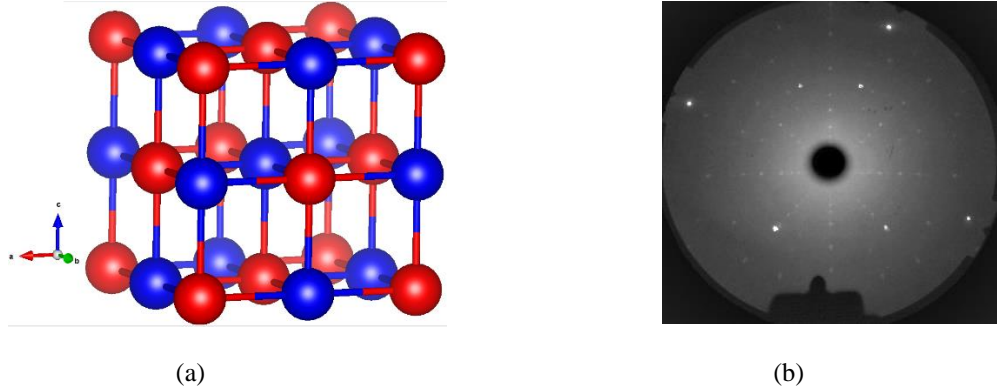

(a)    (b)

**FIGURE 1.** (a) Schematic presentation of the room temperature crystal structure of SmBi. The red and blue colored solid spheres denote the Sm and Bi atom respectively. (b) Laue *x*-ray back diffraction pattern of SmBi (001) single crystal.

The *x*-ray photoemission (XP) spectroscopic measurements on SmBi was carried out using a Phoibos150 analyzer from Specs GmbH using a monochromatic Al $K\alpha$ source. Samples were cleaved *in situ* to expose the clean surface for the measurements. The photoemission measurements were performed at a base pressure of about $2.8 \times 10^{-10}$ Torr. The experiment temperature was achieved using an open cycle helium cryostat, LT-3M from the Advanced Research Systems, USA.

## RESULTS AND DISCUSSIONS

The temperature dependence of the dc magnetic susceptibility was measured in the temperature interval from 5 K - 300 K in the presence of the magnetic field of 40 *k*Oe for zero-field-cooled (ZFC) and field-cooled (FC) conditions using a commercial superconducting quantum interference device (SQUID, Quantum Design, USA). The results are shown in Fig. 2. There is a gradual increase in the magnetic susceptibility, $\chi$ with decreasing temperature down to about 28 K typical for a paramagnetic material. There is a sharp peak at 9 K [see Fig 2(b) where the data are shown in expanded scales] indicating onset of antiferromagnetic ordering, which is in good agreement with the reports of Hulliger *et al.* [14]. These results along with the characterizations described above establish the formation of high quality and stoichiometric single crystals of SmBi.

In order to verify the purity of the sample further, we have taken survey scans using *x*-ray photoemission spectroscopy after cleaving the sample *in situ*. The experimental results show sharp features for various core levels related to Sm and Bi. There was weak intensity (<5% of the other peaks) due to O 1s and C 1s – these features could not be avoided in various trials of cleaving. Despite presence of such small impurities, we did not observe signature of oxides of the constituted elements indicating that the impurity peaks are appearing due to small amount adsorbed oxygens/carbons based compounds on the surface but not strongly bonded to the surface atoms.

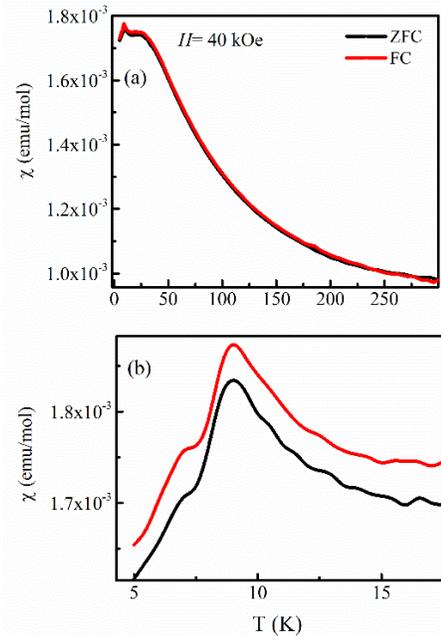

**FIGURE 2.** (a) Temperature dependence of the magnetic susceptibility. (b) Expanded view of the magnetic susceptibility in the temperature range of 4 -18 K.

In Fig. 3, we show the Sm 3d core level spectra obtained using monochromatic Al $K\alpha$ laboratory source at a sample temperature of 140 K at the normal emission geometry [15]. The spectrum in Fig. 3(a) exhibits two sharp peaks at the binding energies of 1081.7 eV and 1109 eV along with two broad features at a binding energy close to 1077.9 eV and 1105.6 eV. The Sm 3d spectra are split into two components due to the spin-orbit interaction having a lower $J$ (= $l - s$) value component, Sm $3d_{3/2}$ and a higher $J$ (= $l + s$) component, Sm $3d_{5/2}$; their energy separation of about 27.3 eV provides an estimate of the Sm 3d spin-orbit splitting. To get more information about these sharp and broad features, we fit the experimental spectrum using asymmetric Gaussian-Lorentzian (GL) product functions. The Gaussian part represents primarily the energy resolution of the instrument and the lifetime broadening of the photoelectrons and the core holes is captured by the Lorentzian width. The fitting was done using least square error method. The simulated spectra provide a good representation of the experimental spectra. In metallic systems, low energy excitations across the Fermi level along with the core hole formation leads to asymmetry of the core level peaks towards higher binding energies, which has been considered for the fittings.

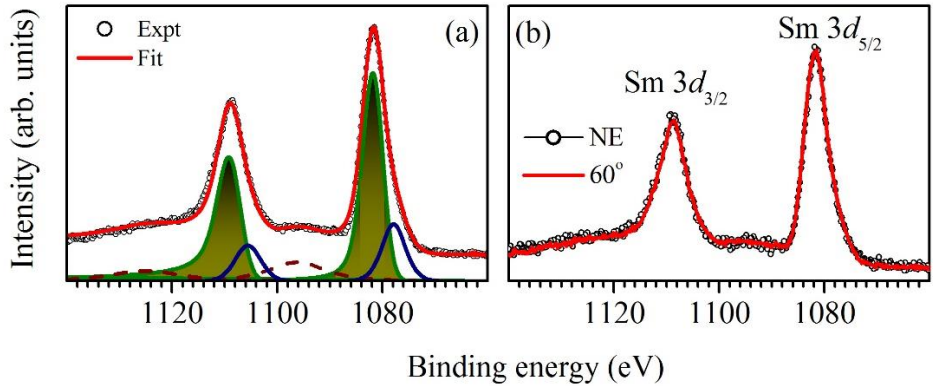

**FIGURE 3.** (a) Core level spectrum of Sm 3d taken at 140 K in normal emission geometry. Experimental data are shown by the open circle and the simulated spectrum is shown by the red line. Solid lines, magenta, and cyan coloured filled plot shows the main peak and the satellite peaks respectively. (b) Normal emission and 60° angled emission.

It is to note here that the experimental spectra exhibit asymmetry in the opposite direction. To capture such unusual spectral shape, it was necessary to consider two peaks. In addition, there are two distinct features at 1096 and 1126 eV. Thus, the simulation of the experimental spectra required three peaks for each of the spin-orbit split photoemission signals. The peaks at 1077.9 eV and 1081.7 eV corresponding to Sm $3d_{5/2}$ signal can be attributed to $4f^6$ and $4f^5$ final states as found in Ce-based compounds [16]. The feature at higher binding energy of 1096 eV presumably appears due to Kondo states formed via Sm 4f – Bi 6p hybridizations. Confirmation of this attribution requires more studies at different temperatures.

In Fig. 3(b), we show the Sm 3d core-level spectrum collected at normal emission and 60° angled emission geometries. There is no significant change in the spectral weight as a function of binding energy, peak position and asymmetry of either the main or satellite peaks. The change in emission angle changes the probing depth of the technique. Here, for Al $K\alpha$ photon energies, the photoelectron kinetic energy is close to 400 eV. Thus, the escape depth is expected to be about 13 Å, which becomes close to 6.5 Å at 60° emission geometry, where the technique is most surface sensitive [17]. Still, we do not observe any change in the spectra indicating almost similar surface and bulk electronic structures. Interestingly, the Kondo feature also does not change suggesting similar surface and bulk Kondo temperature. Confirmation of such conclusions require further studies in this direction.

## CONCLUSIONS

In summary, we have grown high-quality single crystals of SmBi which are predicted to show topological order from density functional theory calculations. Various characterizations established high quality of the single crystals. We have performed magnetic measurements on this crystal exhibiting a sharp antiferromagnetic transition at ~ 9 K. In addition, we have performed XPS measurements on SmBi by varying electron emission angle to probe the surface-bulk differences in the electronic structure. Sm 3$d$ core level spectra exhibit signature of multiple final states indicating finite hybridization of Sm 4f states with the conduction electrons, which is important for Kondo physics. We observe asymmetry in the core-level line shape akin to a metallic system. We further observe fairly intense satellite features associated to the core level excitations presumably a signature of Kondo effect. The core level spectra do not show signature of surface-bulk difference in the electronic structure.

## ACKNOWLEDGMENTS

KM acknowledges financial assistance from the Department of Science and Technology, government of India under J. C. Bose Fellowship program and the Department of Atomic Energy under the DAE-SRC-OI Award program.